# Exploring Design To Environment Methods Though Grassroots Initiatives

Maud Rio[a]*, Benjamin Tyl[b]

[a]*Univ. Grenoble Alpes, CNRS, Grenoble INP, G-SCOP, 38000 Grenoble, France*
[b]*APESA, 40220 Tarnos, France*

* Corresponding author. Tel.: +334-7657-4835; *E-mail address:* maud.rio@g-scop.eu

**Abstract**

Targeting sustainability in our industrial society requires integrating specific criteria in the design process of products and processes. A paradigm shift is necessary in the economical, social and political systems to ensure the natural ecosystems preservation on the planet while fulfilling society needs. Various research methods have therefore emerged to change the way products and services are designed, developed, used and discarded considering territorial contexts. Design for Sustainability, Design for Sustainable Transition, Socially Responsible Design, post-growth design, etc. provide several methods integrating the sustainable principles in the design process. However those approaches remain mainly experimental and are limited to the industrial context. In parallel to those approaches a wide variety of grassroots initiatives have emerged in territories. They propose alternative ways to design systems and they integrate new constraints in a practical manner. This research therefore aims to confront the diversity of Design to Environment (DtE) approaches with 'grassroots' initiatives in order to understand the possible evolution of the integration of sustainability into the design process of products and services used in industry. This paper presents the first literature review results of a started project in 2020. An original research protocol is proposed in this paper, based on specific focus groups with grassroots initiatives practitioners and eco-design experts from research and industry. The presentation of four DtE frameworks are analysed in this paper. This research finally discusses the opportunity of integrating the grassroots enriched DtE frameworks by non-official-designers in life cycle engineering. This bottom-up process may drive an expression of sustainability in industry aligned with some emerging local socio-technical systems.

## 1. Introduction

The complex environmental and social issues we are currently facing require a reinforcement of the approaches towards sustainability that are already engaged in our industrial systems. The socio-political and economic spheres need to be considered together within the planetary boundaries. Hauschild and al. have described the challenges addressed to the Life Cycle Engineering (LCE) within an absolute sustainability consideration. A top-down perspective is combined with a bottom-up perspective coming from the product designers' process [1]. The systemic and multi-impact analysis (eg. LCE methods, parametric, prospective models) to design products and services inside the planetary boundaries need to be developed. The Design for Sustainable transition methods available in the scientific literature are willing to structure such a bottom-up process [2]. Although except from some operational socio-technical singular contexts, or from case studies, Design for Sustainable Transition (DfS) transition processes are mostly remaining theoretical. By contrast some grassroots initiatives propose and demonstrate by their own practice new ways to produce and to consume. Such initiatives are usually included in a territorial approach (eg. the *Atelier Paysan* in the French Alps). The poor consideration of those practices in design methods development illuminates the distinction made between the *conventional designer* (usually an engineer with a business achievements record) way of doing and a rather non-conventional ones (eg. a *Maker, hackers*). These new design practices integrate additional design criteria such as: "local communities' empowerment, proportionality, low intensity, conviviality within small-scale and low-tech solutions" [3]. The design processes are not necessary based on specific design methods. Grassroots initiative actors rather "translate technologies suitable for degrowth societies into goods and services designed for the well-being of communities, through a common approach, and far from profit" [3], as illustrated in section 1.3.

*1.1. Design methods*

Design to Environment (DtE) is defined in this research as a general method merging the Design for X ones contributing to design products and services inside the planetary boundary [4], X stands for eco-efficiency; reuse; repair; remanufacture; etc. DtE methods support designers to integrate the systemic vision needed for an environmental (and sometimes social) consideration in design [5]. The targets to reach sustainability differ from research communities at a conceptual level. In fact product designers have not necessarily opposite visions in the design process they adopt, but rather in the ideal they are willing to contribute to. A generic term of design method is used in this paper to merge all of them.

A recent work of Ceschin and Gaziulusoy [6] identifies levels to characterize the different DtE methods:

- The product level consists of design approaches willing to improve existing products or developing new products.
- The product-service system level supports the development of a combination of products and services.
- The spatio-social level embraces the communities' context of innovation–*conditions*–at different territories scales.
- The socio-technical system level integrates radical changes, focusing on societal needs.

The interaction level of those methods is presented to designers through: (1) the scope of the design intervention and (2) the framing of the design problem, which could be relevant for any grassroot initiative context.

*1.2. Local initiatives*

Local initiatives are supported by several groups of people federated by strong common social values. They take part in developing solution systems to face problems they perceive. Their adopt criteria that usually deviate from traditional institution ones [7]. Contributions are voluntary with a high degree of participation in decision-making. The cause is shared by essence (social value) [8]. Local initiatives enable skills and knowledge development around a given initiative [9]. Observing such initiatives helps to identify the way stakeholders of the civil society are addressing novel discourses–often considered *ahead of the game* [7]. Grassroots initiatives historically focused on local food systems and on local energy production. New design-oriented forms of initiatives have recently emerged in repair workshops, fabLab, low-tech labs for examples.

*1.3. Illustration of design-oriented grassroots initiatives: the low-tech Lab and Atelier Paysan*

The *low-tech Lab* is a French non-profit general interest organization animated by a group of individuals willing to incite people all around the world *to do better with less*. In the manifesto published on their website the term *low-tech* is defined as some *useful, accessible and sustainable* techniques, technologies, services and know-hows. A note from the French foundation Think and Do-tank *la Fabrique Ecologique* strengthens and illustrates the definition given. Their contribution built in a community spirit is composed of Wiki and Tutorials of design projects, a calendar of low-tech activities, and an exchange group *(open to everyone everywhere*, free and accessible online). Some explorations and specific experimentations are also led by this community to investigate in deep some topics. Any voluntary group can be accompanied to create a local low-tech Lab. Official local branches are situated in Grenoble and in Boulogne-Billancourt, linked to an emerging number of local communities (NGO acting around low-techs). The design projects are ranked in prototypes from research process to tutorials, covering various application categories such as water, energy, material, tools. The additional selection criteria proposed include the cost and the difficulty level of the design project shared. Tutorials are referenced, described and illustrated in an explicit manner inviting non-engineers to understand the design steps to follow with supporting videos. Anyone can comment each step to share good practices. Techniques are accessible to non-engineers but are based on engineer know-how. Common mechanical engineering materials and tools are required to develop the projects shared.

Another local initiative illustration providing adapted engineering based toolbox can be founded in the *Atelier Paysan* (cf. website) i.e. 'peasant workshop'. In addition to the Low-tech Lab training courses are given by the cooperative. Their legal structure is similar to the Community Interest Company in the UK, which means that users, employees and partners are sharing the same goal. Their activity is centred on farmer-driven technologies and practices–also accessible, useful and sustainable. This cooperative promotes farmer-driven technical factsheets innovations accessible through an internet forum. In 2020 the team counted three mechanical engineers assisting working groups for compiling specification sheets for the tools to develop, designing and prototyping them (eg.: the *neo-bucher*, the *bed ridger*, the *crosskage roller*). Engineering skills, computer aided design tools, as well as metalworking skills, such as forming or welding, are taught to be replicable. Training courses have been collectively designed to respond to so-called 'real need' felt by the farmers they work with. Debates are therefore occurring to collectively decide whether a technology is worth to be developed to answer a given need or not (eg.: drone-powered or automated weeding systems).

Both illustrations are using, developing, sharing, and challenging design methods to promote a form of technological sovereignty for (or empowerment of) 'non engineers'. Technologies are meant to be resilient, adapted to socio-technical contexts and mastered by users. The organisational, social, societal, commercial, cultural and systemic attributes/components of innovation are explored in such initiatives.

*1.4. Research question*

By contrast to LCE design processes coming from ever increasing sophisticated industrial system and from eco-design experts, some local initiatives are 'popularizing' design processes. They intentionally–or not–use design methods, linked to a territorial context, as well as mobilize their skills and own experiences. In parallel, DtE methods have been widely studied in academic literature, even if there is no evidence that these methods are really applied in industries or by experts.

This research aims at understanding the potential evolution of current DtE methods that integrate sustainability challenges into product design. In other words, the research question can be expressed as follows: which contribution can be brought by local initiatives to design methods adapted to integrate sustainability in the design processes?

This research will then consequently bring some results on this second question: how existing design methods meet (or not) the challenges of current local initiatives? To do so, this research is built upon the following hypothesis: local initiatives and non-conventional designers can bring new perspectives to integrate sustainability into the industrial design processes. In other words, such initiatives (such as repair cafés, fab labs, repair workshops,…), aside from the industrial scheme, may be under-considered in DtE methods.

*1.5. Research project*

The paper originality is to investigate the potential inputs that local initiatives could provide to the existing design methods that support a change of mindset. Therefore, we propose a grassroots initiative-oriented approach.

This paper presents the first results of a started project in 2020 that aims at understanding the original paths to integrate sustainability into the design process of products and services. Design contexts, manufacturing environments, design methods, tools and designers are part of the analysis. This research is based on an experimental confrontation of selected design frameworks to grassroots initiatives practitioners. The result of this research will be to provide to design experts community (such as the CIRP community) a selection of design frameworks enriched with these local initiatives to better integrate sustainable criteria.

Section 2 presents the research method undertaken to address the paper objective. The proposed method is based on a selection and a formalization of recent design frameworks taken from the literature. A protocol based on focus groups to confront the selected frameworks to the local initiative designers is detailed at the end of section 2. Four frameworks are illustrated in section 3 in preparation to the focus group. The last section discusses the method proposed and concludes on the expected results that could drive a shift of paradigm into the current DtE practices.

**2. Research method**

This section details the different steps of the research method adopted.

*2.1. Step 1: The design methods selection*

The first research step aims at identifying and selecting among the DtE frameworks, several associated design methods taken from the scientific literature. The objective is to cover a wide range of DtE methods. Those latter will be adapted to companies design process, and made compatible to local initiatives contexts and singularities.

To do so, the following two criteria were used. First criteria–the identified methods were selected according to two main orientations: (1) methods oriented to a rational point of view (benefic/cost) vs. methods oriented towards a qualitative and social point of view; (2) methods oriented towards a limits/constraints point view vs. methods oriented towards a visionary point of view. Second criteria–the identified methods were selected according to their degree of formalism and in accordance to their 'engineering format' compatibility. They are easy to understand, presented in an explicit manner, and documented in detail in the literature.

*2.2. Step 2: a table formalizing the selected frameworks*

The second step consists of describing the DtE methods selected according to the following analysis grid (*cf.* Table 1).
- The framework: refers to a way of thinking - a paradigm - shared by each DtE methods included in that framework.

As exemplified in section 3, *sustainable transition*, or *business models* are defined in two distinct design frameworks. A design method supporting a 'business model' will therefore be understood by referring to the rational system of thinking.
- The key principle is the ideology underlying the framework.

For example, the key principle for business models is to analyze systems according to the value proposition, to the value delivery, and to the value capture.
- The support is the framework formalization into a practical method.

Examples of supports are guidelines, canvas, toolbox, etc. Each design methods should finally be scientifically referenced.

*2.3. Step 3: a protocol to confront the selected frameworks to the local initiative designers*

The step 3 consists in developing several focus groups to confront the different design frameworks to the local initiatives' vision referring to the methodology brief published in [10]. Focus groups are defined in this research as discussion groups "organized to explore a particular set of issues" [11]. This confrontation has the following objectives: to explore and collect the local initiative expertise, his/her beliefs and opinions about the provided design frameworks used in industry. This process operates using the *own* language and vocabulary of the focus group participants. Misunderstandings are kept minimal by avoiding implicit meanings in expressions, using reformulation technics. These focus groups are oriented on the following subjects and objectives:

- Subject 1: how the eco-design methods presented meet the challenges of current local initiatives?
- Objective 1: determine the relevance of these design methods for local initiatives; characterize the functionalities provided that covers the local initiatives'

- supports lacks (eg: bring a systemic vision currently lacking by the available methods).
- Subject 2: what are the shortcomings of these methods within local initiatives?
- Objective 2: identify some opportunities for current design methods improvements to embrace the grassroots initiative's needs and singularities (to be expressed in an explicit manner).

Three focus groups are being constituted respectively in Grenoble, Troyes (or Paris), and Bordeaux. The Grenoble participants have been already selected and invited, chosen through similar characteristics (cf. [10] about the focus group methodology): local initiative designers, practicing a wide range of mechanical engineering technics (from traditional ones to additive manufacturing ones used in FabLabs). Their have a role of design manager, supervising and/or designing, fabricating, or repairing, a product or a system, in a FabLab, a repair shop (etc.) context. They share the same characteristics with different contexts previously visited. Several design project examples are taken from their context to prepare some illustrations of the application of the selected DtE frameworks. A design process example is taken for instance from "La Bonne Fabrique" place, where some craft beers are brewed. The paper based labels are made from the process distiller's grains, during local workshops, cut and printed with a stamp made with the laser cuter from the FabLab, stuck with local milk (casein is used as a glue).

A set of feedbacks formulated by each local initiative group on the same DtE methods frameworks will be cross-analyzed, and presented during the research method's fourth step. Such experiment requires a common DtE method presentation given to each focus group. Section 3 transcribes four of the selected frameworks to deliver the same explanation to each focus group.

*2.4. Step 4 : a protocol to confront the results of step 3 with eco-design experts*

The last step of this research method consists in presenting and discussing the feedbacks from local initiatives with eco-design experts from the EcoSD French research network. This network gathers experts from industries, consulting, research and public agencies that will constitute the last focus group, oriented on the following subject and objective:
- Subject 3: how cross-analyzed feedbacks from local initiatives may support an eco-design processes evolution in industry?
- Objective 3: identify the opportunities of eco-design practices developments in life cycle engineering.

**3. First results: selection and analysis of DtE frameworks, explanations given to focus groups**

Four main frameworks are considered in this research and described in this section. Each description is associated to a graphical representation confronted to the focus group participants. Table 2 summarizes the selected DtE methods according to the analysis grid exposed in section 2.2. This section describes the explanation given to the participants' focus groups.

Table 1. DtE method analysis addressed to local initiatives

| Frameworks | Principles | Support | Ref. |
|---|---|---|---|
| Conviviality *Limits oriented framework* | Identification of 6 threats society must face in order to avoid over-industrialization | Matrix & guidelines | [14,15] |
| Business models *Rational framework* | Analysis of the systems according to the value proposition, value delivery and value capture | Sustainable business models archetype Triple layered canvas | [17,18] |
| Social design *Qualitative and social oriented framework* | Integrating social and unquantifiable issues in the design process (health, local, empowerment …) | Socially responsible design Design for Sustainability (UNEP) | [25,26] |
| Sustainable transition *Visionary framework* | Imagine a future system within sustainability objectives | Double flows method Framework for Strategic Sust. Dev. (FSSD) Five dimension-sustainability | [22,27,28] |

*3.1. The conviviality framework*

The concept of conviviality is about living in accordance with a system that satisfies human needs through the contributions of autonomous individuals, rather than with the principles of industrial society [12]. *Conviviality* therefore designates the opposite of industrial productivity. The *tools for conviviality* for Illich are easy to use if required (or not) on purpose determined by the users themselves. Using such a tool should not infringe someone else freedom in using it as well. No specific qualification is required to use a convivial tool. The systemic approach is central in conviviality ensuring the relationship between individuals, their environment and their technology [13]. A *convivial society* is defined by Illich through six threats the society must face in order to avoid over-industrialization: (1) biological degradation (i.e. over use of resources, energies, crossing planetary boundaries); (2) radical monopoly (i.e. the Human dependence on industrial products); (3) over-programming (i.e. tools conferring an intentional and programmed training rather than focusing on human creativity); (4) polarization (i.e. a society that inevitably concentrates power in the hands of a few people deciding the future of everyone); (5) obsolescence (.i.e technical obsolescence of tools, in addition to the associated known-how obsolescence); and (6) the frustration caused by several threats operating simultaneously. Voinea (2017) adapted the Illich's approaches through two axes and six criteria: (1) the personal autonomy, such as the flexibility, transparency, simplified and usability; and the social cohesion, such as the 'sharedness', the creativity, and the sociality. Vetter developed a design tool - the matrix of convivial technology to assess technologies [14], whereas Lizarralde and Tyl proposed some integrated design guidelines for conviviality [15].

*3.2. Sustainable Business Models*

The business model innovation is widely studied in the literature, and recognized as a key to reach sustainability [16,17]. The approach basically consists in explaining the (sustainable) value, the way the value is delivered, and users that are targeted. In the sustainability field, Bocken et al. [18] identified 8 archetypes of sustainable business models from a technological, social and environmental point of view: (1) maximizing material productivity and energy efficiency; (2) creating value from 'waste'; (3) substituting with renewables and natural processes; (4) delivering functionality, rather than ownership; (5) adopting a stewardship role; (6) encouraging sufficiency; (7) re-purposing the business for the society/environment; (8) developing scale-up solutions. Joyce and Paquin [18] adapted the business model canvas in a practical way [19] by underlining the environmental and social issues through a *Triple Layered Business Model Canvas (TLBMC)*. The canvas environmental issue is based on a life cycle principles, whereas the social issues are based on the stakeholders' point of views in regard to his/her environmental impact influence.

*3.3. Transition design framework*

The transition design focuses on a system analysis and on the evolution of this system in regard to the social, institutional and technological changes that could be enabled in the future [20]. This systemic approach focuses on some fundamental changes "in the ways of organizing (structures), ways of thinking (cultures) and ways of doing (practices)" [21]. The approach consist in imagine the future system within some sustainability objectives; exploring the different pathways to reach such future systems; translating them into practical actions; and assessing the potential success reached [21]. To support such a framework, Gaziulusoy et al. [22] proposed a systemic double-flow design scenario method to support companies in reaching sustainability within a systemic approach. The method stands on a strong sustainability approach. Medium to long-term objectives are encouraged to help companies identifying the technology and organizational developments required. A global business strategy is then required to ensure the operationalization of the objectives in a societal vision aligned with present and future realistic developments. The double flow analysis method consists in three stages: (1) a preparation stage to understand the current systems, the associate risks and the relationships between the environment, the society and the economic spheres; the current system social functions are also identified; (2) a scenario development stage, elaborated within the risks and the realistic vision identified in stage 1. Designers are encouraged to develop a roadmap through forecasting and backcasting approaches. This roadmap is then completed with the current and future stakeholders of the system. A product and service vision completes the roadmap; (3) a final action-plan stage details the different steps to reach the targeted objective.

*3.4. Social design framework*

Social design is probably the less structured framework chosen in this research. As McMahon and Bhamra's literature review described: the social aspects of design are 'quantifiable' elements of sustainability (environmental quantification is assumed) [23]. The proposal of Lilley is added in the author analysis [24] including "personal responsibility, equitable distribution of social capital, meeting basic needs, quality of life, health, well-being and happiness, democratic participation, trusting, harmonious and cooperative behavior and preserving social and cultural dynamism". Social design is therefore often considered with a partial view, also found in frugal innovation, or Design for Bottom of the Pyramid, and socially responsible design. These latter consider social design as "Design 'solutions' range from products or systems that utilize existing or new skills and workmanship, utilize natural local resources and materials and are wholly manufacturable and maintained by the end user communities or nearby regional centers, to 'band aid' or 'parachute' products that are reliant on first world manufacturing and supply chains." Crul et al. [25] developed a practical approach integrating strong social aspects in design for the United Nation Environmental Program: distributed economy, social responsibility, and Human resources management. Melles et al. [26] developed the following criteria to support social-oriented design: need, suitability, relative affordability, advancement, local control, usability, empowerment, and dependency.

**4. Discussion and conclusion**

This research investigates the potential of grassroots initiatives in improving the Design to Environment process in industry. This paper exposes a research protocol to capture this potential in a bottom-up perspective. The first steps 1-3 aims through the focus groups currently conducted at highlighting the strengths and weaknesses of the current DtE methods according to the challenges of current grassroots initiatives locally identified. The step 4 of the protocol aims at identifying the opportunities for eco-design developments. So far, the results published in this paper are based on a systematic literature review. The LCE product development stages will be then enriched by the grassroots initiative contributions in DtE methods. For instance the conviviality design guidelines are rather adapted to the early design stages. Addressing engineers and managers a sustainable business method improved by focus groups feedbacks at the planning and task clarification of the design process would support the value definition engaged in the project. The conviviality design guidelines for socio-technical scope are applicable as early as the product is getting conceptualized. Local initiatives feedbacks from their own design practices may help engineers with prospections (eg. upscalling technology). The TLBMC would be useful at this stage to support the company (or any corporate) to define the business model associated to the product life cycle vision. As soon as the embodiment design stage and the prototyping stage are visited, the 5D method can be used to determine the medium to long-term business strategy of the company. The production and market launch of the product opens to prospection methods. The double flows method for sustainable transition can therefore be useful for steadily moving forward sustainability. At any time during the product reviews the product designers can evaluate the dimension of the convivial technology potential that has been developed. The matrix of convivial technology can be used for a self-assessment of degrowth oriented organizations developing or adapting technologies. As argued by the author this method is a way to challenge the social imaginary concerning a given technology that would be used by the product designers during the project development, and/or, promoted by the organization (local initiative, industry).

In conclusion, the originality of this research is to investigate the grassroots initiatives design processes and organizational contexts to enrich current design frameworks. Non-conventional designers have a potential to bring new perspectives for integrating sustainability into industrial processes. The experimental stage started in early 2021 with the focus groups challenges the selection of design methods identified in this paper, based on a systematic literature review as relevant and promising to support bottom-up strategies improving life cycle engineering DtE methods.

This research protocol will be used to explore design frameworks in other research projects as well. A similar protocol may be adapted to the specific sector of repair workshops, to analyze the methods supporting the development of repair and reuse ecosystems. In this case the focus groups will gather experts and repair workshops practitioners (TERROIR project).

## Acknowledgements

The authors want to thank the EcoSD network and ADEME for funding this Collaborative Research Project called *SustainLives*. They also acknowledge the support of ADEME (TEES program) for funding the TERROIR project as well as the region Nouvelle Aquitaine.

## References

[1] Hauschild M. Z., Kara S., Ropke I. (2020). Absolute sustainability: Challenges to life cycle engineering. CIRP Annals – Manufacturing Technology 69, 533-553.
[2] Gaziulusoy, I., Erdoğan Öztekin, E. (2019). Design for sustainability transitions: Origins, attitudes and future directions. Sustainability, 11(13), 3601.
[3] Tyl, B. Lizarralde, I., Vetter, A. (2018). The design approach within a degrowth perspective, special call of paper, 6th Degrowth conference, Malmo.
[4] Steffen, W., Richardson, K., Rockström, J., Cornell, S. E., Fetzer, I., Bennett, E. M., ... & Folke, C. (2015). Planetary boundaries: Guiding human development on a changing planet. Science, 347(6223).
[5] Rio, M., Riel, A., Brissaud, D. (2017). Design to environment: information model characteristics. Procedia CIRP, 60(1), 494-499.
[6] Ceschin, F., Gaziulusoy, I. (2016). Evolution of design for sustainability: From product design to design for system innovations and transitions. Design studies, 47, 118-163.
[7] Gernert, M., El Bilali, H., Strassner, C. (2018). Grassroots initiatives as sustainability transition pioneers: implications and lessons for urban food systems. Urban Science, 2(1), 23.
[8] Grabs, J., Langen, N., Maschkowski, G., Schäpke, N. (2016). Understanding role models for change: a multilevel analysis of success factors of grassroots initiatives for sustainable consumption. Journal of Cleaner Production, 134, 98-111.
[9] Smith, A.; Stirling, A. (2016). Grassroots Innovation and Innovation Democracy; STEPS, Working Paper 89; STEPS Centre: Brighton, UK.
[10] Grudens-Schuck, N., Allen, B. L., Larson, K. (2004). Methodology Brief, Focus Group Fundamentals. Extension Community and Economic Development Publications, 12.
[11] Holloway, I. (2005). Qualitative research in health care. Berkshire, UK: Open University Press.
[12] Illich, I., (1973). Tools for Conviviality. Calder & Boyars, London.
[13] Voinea, C. (2018). Designing for conviviality. Technology In Society, 52, 70-78.
[14] Vetter, A. (2018). The matrix of convivial technology–assessing technologies for degrowth. Journal of cleaner production, 197, 1778-1786.
[15] Lizarralde, I., Tyl, B. (2018). A framework for the integration of the conviviality concept in the design process. Journal of Cleaner Production, 197, 1766-1777.
[16] Lüdeke-Freund, F. (2010). Towards a conceptual framework of' business models for sustainability'. Knowledge collaboration & learning for sustainable innovation, R. Wever, J. Quist, A. Tukker, J. Woudstra, F. Boons, N. Beute, eds., Delft.
[17] Bocken, N. M., Short, S. W., Rana, P., Evans, S. (2014). A literature and practice review to develop sustainable business model archetypes. Journal of cleaner production, 65, 42-56.
[18] Joyce, A., Paquin, R. L. (2016). The triple layered business model canvas: A tool to design more sustainable business models. Journal of cleaner production, 135, 1474-1486.
[19] Osterwalder, A., Pigneur, Y. (2010). Business model generation: a handbook for visionaries, game changers, and challengers. John Wiley & Sons.
[20] Loorbach, D., Frantzeskaki, N., Avelino, F. (2017). Sustainability transitions research: transforming science and practice for societal change. Annual Review of Environment and Resources, 42.
[21] Gorissen, L., Vrancken, K., Manshoven, S. (2016). Transition thinking and business model innovation–towards a transformative business model and new role for the reuse centers of Limburg, Belgium. Sustainability, 8(2), 112.
[22] Gaziulusoy, A. I., Boyle, C., McDowall, R. (2013). System innovation for sustainability: a systemic double-flow scenario method for companies. Journal of Cleaner Production, 45, 104-116.
[23] McMahon, M., Bhamra, T. (2015). Social sustainability in design: Moving the discussions forward. The Design Journal, 18(3), 367-391.
[24] Lilley, D. (2007). Designing for behavioural change: reducing the social impacts of product use through design. Doc. Thesis, Loughborough Univ.
[25] Crul, M., Diehl, J. C., Ryan, C. (2006). Design for sustainability. A practical approach for developing economies, United Nation Environmental Programme, TU Delft, Paris.
[26] Melles, G., de Vere, I., Misic, V. (2011). Socially responsible design: thinking beyond the triple bottom line to socially responsive and sustainable product design. CoDesign, 7(3-4), 143-154.
[27] Broman, G. I., Robèrt, K. H. (2017). A framework for strategic sustainable development. Journal of Cleaner Production, 140, 17-31.
[28] Allais R. (2015). Transition systémique pour un développement durable: entre conception et territoire. Doc. Thesis, UTT